# Distributed Recursive Filtering for Spatially Interconnected Systems with Randomly Occurred Missing Measurements

Bai Li[1]

[1] Department of Control Science and Engineering

Zhejiang University

No.38 Zheda Road, Hangzhou, China

libai@zju.edu.cn

ABSTRACT. *This paper proposed a distributed filter for spatially interconnected systems (SISs), which considers missing measurements in the sensors of sub-systems. An SIS is established by many similar sub-systems that directly interact or communicate with connective neighbors. Despite that the interactions are simple and tractable, the overall SIS can perform rich and complex behaviors. In actual projects, sensors of sub-systems in a sensor network may break down sometimes, which causes parts of the measurements unavailable unexpectedly. In this work, distributed characteristics of SISs are described by Andrea model and the losses of measurements are assumed to occur with known probabilities. Experimental results confirm that, this filtering method can be effectively employed for the state estimation of SISs, when missing measurements occur.*

1. **Introduction.** In recent years, models of spatially interconnected systems (SISs) have been widely adopted to describe large scale systems or temporal-spatial systems [1,2], in which many similar units directly interact with their connective neighbors. It is confirmed that despite such units interact with their neighbors in a simple and tractable way, the overall lumped system is able to perform rich and complex behaviors [3]. Applications of SIS are in large quantities, for instance, intelligent vehicle highway systems [4], autonomous flight formation systems [5], and even financial market systems [6].

Filtering is an essential process for SISs. Filters definitely work to estimate intrinsic signals by means of denoising or orthogonalizing. One feasible filtering method for SISs is the $H_\infty$ model [7]. In $H_\infty$ methods, at each sampling time, the necessary Linear Matrix Inequality (LMI) condition of recursion needs to be verified. In other words, to implement $H_\infty$ filtering algorithms recursively is not easy. Andrea model is another method on the basis of state space, which satisfies LMI condition more easily [3]. This work introduces Andrea model to describe SISs.

When distributing over a wide area, sensors in a network may inevitably suffer from significant communication delays or data losses. In a way, significant delay is equivalent to loss, because only timely derived data is worthy to the subsequent control process. Researches on state estimation with uncertain observations begin from Nahi Nasser in 1969 [8]. Nahi introduced the formula of Bernoulli binomial distribution to describe the occurrence of observation failure and assumed that some observations contain noises alone. Reference [9] proposed a robust Kalman filtering approach considering both observation delays and losses. Reference [10] introduced a recursive filtering algorithm for nonlinear systems with missing measurements. But note they are not suitable for distributed systems. Regarded with SISs, system observations with constant bias errors [11] and with quantization errors and successive packet dropouts [12] have been settled. But these filters cannot implement recursively. To briefly conclude, researchers have seldom developed recursive filters with missing measurements for distributed systems [13].

In this work, measurement losses of sensors are assumed to be randomly unavailable with prior known probabilities. I propose a two-step filtering algorithm, in which the interconnection variables among sub-systems are estimated in the first step, and then local state estimation is implemented recursively for each sub-system, taking missing measurements considered.

The remainder of this paper is organized as follows. In Section 2, basic principle of the Andrea model is introduced, and the statistical description of measurement losses is also given as a necessary background. Procedure of the two-step filtering algorithm is elaborated in Section 3. And then some concluding remarks are drawn in the last section.

2. **Description of SIS with missing measurements.** This section introduces a slightly modified Andrea model at first, and then describes the missing measurements in Bernoulli sequences. Fig. 1(a) shows a schematic diagram of SIS $\{\Sigma\}$ together with local sub-filters $\{F\}$, where $\hat{x}(t,i)$ denotes the $i$th estimated state at time $t$, and $u(t,i)$ denotes the given input for the $i$th sub-system. It is supposed that $S_m$ is finite here and this work merely focused on SIS of one spatial dimension. Note that each local filter $F(i)$ can utilize only observation signals from its corresponding local sub-system and the adjacent signals from its nearest neighbor(s). Fig. 1(b) depicts a detailed structure of each sub-filter, $w(t,i)$ and $v(t,i)$ respectively represent interconnected signals sent in and out of the $i$th sub-filter. Linear time-variant subsystems can be established by the following state-space equations.

$$x(t+1,i) = A \cdot x(t,i) + B \cdot v(t,i) + C \cdot u(t,i). \tag{1}$$

$$w(t,i) = D \cdot x(t,i) + G \cdot v(t,i) + H \cdot u(t,i). \tag{2}$$

It should be noted that matrices $A$, $B$, $C$, $D$, $G$, and $H$ are generally functions of $t$ and $i$, these arguments are dropped here for simplicity in notation. The interconnected variables (see Fig. 1(b)) can be defined as follows.

$$v(t,i) = \begin{bmatrix} v_+(t,i) \\ v_-(t,i) \end{bmatrix}, \quad w(t,i) = \begin{bmatrix} w_+(t,i) \\ w_-(t,i) \end{bmatrix}, \quad \begin{cases} v_+(t,i) = w_+(t,i-1) \\ v_-(t,i-1) = w_-(t,i) \end{cases}, \quad i \in [2, S_m] \cap \Box. \tag{3}$$

Since $S_m$ is not infinite, hence the boundary conditions are considered as well:

$$v_+(t,1) \equiv \mathbf{0}, \quad v_-(t,S_m) \equiv \mathbf{0}, \quad w_-(t,1) \equiv \mathbf{0}, \quad w_+(t,S_m) \equiv \mathbf{0}. \tag{4}$$

Besides, observations do not necessarily contain states all the time. Losses in the observations can be described by the scalar $\gamma(t,i)$ as in (5), which is assigned to be 1 or 0. That is to say, $p(t,i)$ denotes the probability that observation of the $i$th subsystem contains state $x(t,i)$ at time $t$. $d(t,i)$ represents a zero-mean Gaussian noise with covariance matrix $M$, which is added to the observation signal.

$$y(t,i) = \gamma(t,i) \cdot J \cdot x(t,i) + R \cdot u(t,i) + d(t,i), \quad (5)$$

with

$$\begin{cases} p(t,i) \triangleq \operatorname{Prob}(\gamma(t,i)=1) \\ 1-p(t,i) \triangleq \operatorname{Prob}(\gamma(t,i)=0) \end{cases}.$$

With the above described model, this work focuses on recursively estimating the state $\hat{x}(t,i)$ using randomly missing observation $y(t,i)$ and the given input signal $u(t,i)$ in each sub-system.

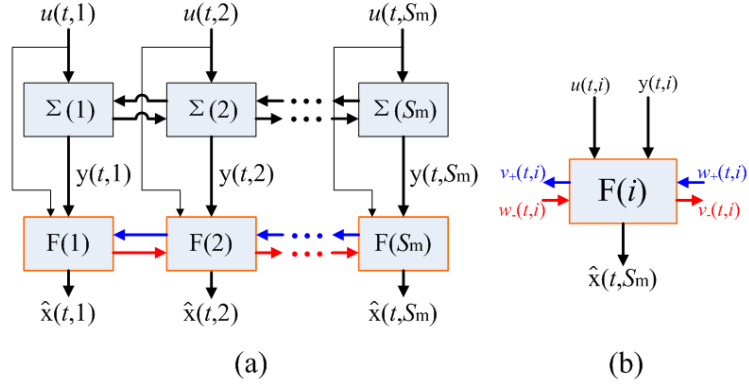

(a)　　　　　　　　　　　　(b)

FIGURE 1. Schematic diagram of Andrea model in one dimension; (a) Overall diagram; (b) Local diagram of sub-filter

3. **Derivation of recursive distributed filter for SIS.** Regarding system (1,5), it is confirmed that each $\hat{x}(t+1,i)$ can be predicted and updated by standard Kalman filter, on condition that each $u(t,i)$, $v(t,i)$, and $y(t,i)$ are given or obtained [8]. On the other hand, $v(t,i)$ can be obtained using the relationship between $w$ and $v$, provided that $u(t,i)$ and $x(t,i)$ are known. When current measurements involve losses, a modified Kalman filter as in [8] is introduced to substitute for the standard Kalman filter. In brief, this section proposed a recursive filter in two steps, one is to obtain the interconnected vector $v(t,i)$ on the basis of known $u(t,i)$ and $x(t,i)$; the other is to estimate each local state $\hat{x}(t+1,i)$ concerning imperfect observations, which drop internal true states with prior known probabilities.

First, it is assumed that $x(t,i)$ and $u(t,i)$ are already known, then $v(t,i)$ is expected to derive. Let a new variable $z(t,i)$ replace $D \cdot x(t,i) + H \cdot u(t,i)$ to describe the known part as in (1). Then (6) is derived.

$$\begin{bmatrix} w_+(t,i) \\ w_-(t,i) \end{bmatrix} = \begin{bmatrix} G_{11} & G_{12} \\ G_{21} & G_{22} \end{bmatrix} \cdot \begin{bmatrix} v_+(t,i) \\ v_-(t,i) \end{bmatrix} + \begin{bmatrix} z_+(t,i) \\ z_-(t,i) \end{bmatrix}, \quad (6)$$

where $G(t,i)$ and $z(t,i)$ are partitioned into submatrixes in accordance with the sizes of $w_+(t,i)$ and $w_-(t,i)$. If $G_{22}(t,i)$ is invertible, (6) is rewritten in a lumped form:

$$\begin{bmatrix} w_+(t,i-1) \\ v_-(t,i) \end{bmatrix} = \begin{bmatrix} I & O \\ -G_{22}^{-1}G_{21} & G_{22}^{-1} \end{bmatrix} \cdot \begin{bmatrix} w_+(t,i-1) \\ v_-(t,i-1) \end{bmatrix} + \begin{bmatrix} O & O \\ O & -G_{22}^{-1} \end{bmatrix} \begin{bmatrix} z_+(t,i) \\ z_-(t,i) \end{bmatrix}, \quad (7)$$

$$\begin{bmatrix} w_+(t,i) \\ v_-(t,i) \end{bmatrix} = \begin{bmatrix} G_{11} & G_{12} \\ O & I \end{bmatrix} \cdot \begin{bmatrix} w_+(t,i-1) \\ v_-(t,i) \end{bmatrix} + \begin{bmatrix} I & O \\ O & O \end{bmatrix} \begin{bmatrix} z_+(t,i) \\ z_-(t,i) \end{bmatrix}. \quad (8)$$

Substituting (7) into (8) yields

$$\begin{bmatrix} w_+(t,i) \\ v_-(t,i) \end{bmatrix} = \begin{bmatrix} G_{11} - G_{12} \cdot G_{22}^{-1} \cdot G_{21} & G_{12} \cdot G_{22}^{-1} \\ -G_{22}^{-1} \cdot G_{21} & G_{22}^{-1} \end{bmatrix} \cdot \begin{bmatrix} w_+(t,i-1) \\ v_-(t,i-1) \end{bmatrix} + \begin{bmatrix} I & -G_{12} \cdot G_{22}^{-1} \\ O & -G_{22}^{-1} \end{bmatrix} \begin{bmatrix} z_+(t,i) \\ z_-(t,i) \end{bmatrix},$$

which can be rewritten as follows to highlight the recursive characters:

$$\begin{bmatrix} w_+(t,i) \\ v_-(t,i) \end{bmatrix} = \phi(t,i) \cdot \begin{bmatrix} w_+(t,i-1) \\ v_-(t,i-1) \end{bmatrix} + \beta(t,i), \tag{9}$$

where

$$\begin{bmatrix} G_{11} - G_{12} \cdot G_{22}^{-1} \cdot G_{21} & G_{12} \cdot G_{22}^{-1} \\ -G_{22}^{-1} \cdot G_{21} & G_{22}^{-1} \end{bmatrix} = \phi(t,i), \quad \begin{bmatrix} I & -G_{12} \cdot G_{22}^{-1} \\ O & -G_{22}^{-1} \end{bmatrix} \begin{bmatrix} z_+(t,i) \\ z_-(t,i) \end{bmatrix} = \beta(t,i).$$

In this way, items in the series of (9) can be derived. Eq. (10) refers to $i = 2$, and (11) refers to $i = S_m - 1$.

$$\begin{bmatrix} w_+(t,2) \\ v_-(t,2) \end{bmatrix} = \phi(t,2) \cdot \begin{bmatrix} w_+(t,1) \\ v_-(t,1) \end{bmatrix} + \beta(t,2). \tag{10}$$

$$\begin{bmatrix} w_+(t,s_m-1) \\ v_-(t,s_m-1) \end{bmatrix} = \phi(t,s_m-1) \cdot \begin{bmatrix} w_+(t,s_m-2) \\ v_-(t,s_m-2) \end{bmatrix} + \beta(t,s_m-1). \tag{11}$$

Iterating these items yields

$$\begin{bmatrix} w_+(t,s_m-1) \\ v_-(t,s_m-1) \end{bmatrix} = [\prod_{i=2}^{s_m-1} \phi(t,i)] \cdot \begin{bmatrix} w_+(t,1) \\ v_-(t,1) \end{bmatrix} + \sum_{i=2}^{s_m-2} \{[\prod_{j=i+1}^{s_m-1} \phi(t,j)] \cdot \beta(t,i)\} + \beta(t,s_m-1). \tag{12}$$

Therefore, the general formula for the $i$th sub-system ($i \in [2, S_m - 1] \cap \Box$) is derived in (12). But the initial conditions remain to be determined. Substituting for $i = 1$ into (8) and recalling that $v_+(t,1) = \mathbf{0}$ yields

$$w_+(t,1) = G_{12} \cdot v_-(t,1) + z_+(t,1). \tag{13}$$

Let $\mathrm{MT}(t) = \prod_{i=2}^{s_m-1} \phi(t,i)$ and $\mathrm{MC}(t) = \sum_{i=2}^{s_m-2} \{[\prod_{j=i+1}^{s_m-1} \phi(t,j)] \cdot \beta(t,i)\} + \beta(t,s_m-1)$ for simplicity, and then substituting for (12) yields

$$\begin{bmatrix} w_+(t,s_m-1) \\ v_-(t,s_m-1) \end{bmatrix} = \begin{bmatrix} \mathrm{MT}_{11} & \mathrm{MT}_{12} \\ \mathrm{MT}_{21} & \mathrm{MT}_{22} \end{bmatrix} \cdot \begin{bmatrix} G_{12} \cdot v_-(t,1) + z_+(t,1) \\ v_-(t,1) \end{bmatrix} + \mathrm{MC}, \tag{14}$$

where MT is partitioned according to the sizes of $v_+$ and $v_-$.

Similarly, substituting for $i = S_m$ into (7), and then by substitution again regarding (3), then we have

$$\begin{bmatrix} O & O \\ G_{21} & -I \end{bmatrix} \begin{bmatrix} w_+(t,s_m-1) \\ v_-(t,s_m-1) \end{bmatrix} + \begin{bmatrix} O & O \\ O & I \end{bmatrix} \begin{bmatrix} z_+(t,s_m) \\ z_-(t,s_m) \end{bmatrix} = \mathbf{0}. \tag{15}$$

Substituting (14) into (15) yields

$$v_-(t,1) = [G_{21}(t,s_m) \cdot \mathrm{MT}_{11} \cdot G_{12}(t,1) - \mathrm{MT}_{21} \cdot G_{12}(t,1) + G_{21}(t,s_m) \cdot \mathrm{MT}_{12} - \mathrm{MT}_{22}]^{-1} \cdot \\ [-G_{21}(t,s_m), I] \cdot \mathrm{MC}. \tag{16}$$

Recall the relationship between $w_+(t,1)$ and $v_-(t,1)$ as in (13), we derive the first item:

$$\begin{bmatrix} w_+(t,1) \\ v_-(t,1) \end{bmatrix} = \begin{bmatrix} G_{12}(t,1)\big[G_{21}(t,s_m)MT_{11}G_{12}(t,1) - MT_{21}G_{12}(t,1) + G_{21}(t,s_m)MT_{12} - MT_{22}\big]^{-1} \times \\ \big[G_{21}(t,s_m)MT_{11}G_{12}(t,1) - MT_{21}G_{12}(t,1) + G_{21}(t,s_m)MT_{12} - MT_{22}\big]^{-1} \times \\ \times \big[-G_{21}(t,s_m), I\big] \cdot MC + z_+(t,1) \\ \times \big[-G_{21}(t,s_m), I\big] \cdot MC \end{bmatrix} \quad (17)$$

In this way, items from the first until the $(s_m-1)$th one can be recursively derived as follows.

$$\begin{bmatrix} w_+(t,k) \\ v_-(t,k) \end{bmatrix} = [\prod_{i=2}^{k}\phi(t,i)] \cdot \begin{bmatrix} w_+(t,1) \\ v_-(t,1) \end{bmatrix} + \sum_{i=1}^{k-1}\{[\prod_{j=i+1}^{k}\phi(t,j)] \cdot \beta(t,i)\} + \beta(t,k). \quad (18)$$

Recall that each $w_+(t,i-1)$ equals $v_+(t,i)$ as in (3), so we have

$$v(t,k) = \begin{bmatrix} w_+(t,k-1) \\ v_-(t,k) \end{bmatrix}, \ (k=2,3,\cdots,s_m). \quad (19)$$

In this way, $v(t,i)$ is derived on the basis of (17) and (19) to conclude the first step. In the second step, we assume that $v(t,i)$ is derived, $u(t,i)$, $y(t,i)$, and $p(t,i)$ are given, then $\hat{x}(t+1,i)$ is expected to derive. Let us define two new variables to represent known parts in (1) and (5):

$$c_1 = B \cdot v(t,i) + C \cdot u(t,i), \ c_2 = R \cdot u(t,i). \quad (20)$$

Here, the conclusion in [8] is directly used here to determine the form of each local sub-filter, as in (21).

$$\hat{x}(t+1,i) = \xi_1(t,i) \cdot \hat{x}(t,i) + \xi_2(t,i) \cdot [y(t,i) - c_2] + c_1, \quad (21)$$

where

$$\xi_1(t,i) = A - p(t,i) \cdot \xi_2(t,i) \cdot J,$$

$$\xi_2(t,i) = p(t,i) \cdot A \cdot T(t,i) \cdot J^T \cdot \big[M + p^2(t,i) \cdot J \cdot T(t,i) \cdot J^T + (p(t,i) - p^2(t,i)) \cdot J \cdot S(t,i) \cdot J^T\big]^{-1},$$

$$S(t+1,i) = A \cdot S(t,i) \cdot A^T,$$

$$T(t+1,i) = [A - p(t,i) \cdot \xi_2(t,i) \cdot J] \cdot T(t,i) \cdot A^T,$$

$$S(1,i) = T(1,i) = E\big(x(1,i) \cdot x^T(1,i)\big).$$

Due to the fact that the local filter originally proposed by Nahi in 1963 has been confirmed to be optimal, and that our first step only involves some iterative computation, therefore, there is no need to conduct arbitrarily designed numerical simulations.

4. **Conclusions.** This paper proposed a novel recursive distributed filter for SISs on the basis of state-space framework. Despite the fact that investigations concerning this field mainly focus on $H_\infty$ methods or partial differential equation methods, this work opens up a new way. Applying this proposed approach to settle some practical problems regarding macro-economic control systems and flight control systems will be my future work.

**Acknowledgment.** This work was sponsored in part by the 6[th] National College Students' Innovative and Entrepreneurial Training Programs and supported in part by Professor Yan Lin and Hua-yong Liang.